# On the difference between stiff and soft membranes: Capillary Waves

Sebastian Jaksch[1,*], Olaf Holderer[1], Michael Ohl[2], Henrich Frielinghaus[1]

[1] Forschungszentrum Jülich GmbH, JCNS at Heinz Maier-Leibnitz Zentrum, Lichtenberstraße 1, 85747 Garching, Germany

[2] Forschungszentrum Jülich GmbH, JCNS at SNS-Oak Ridge National Laboratory (ORNL), 1 Bethel Valley Road, Oak Ridge, TN 37831, USA

## Abstract

One problem of non-crystalline condensed matter (soft matter) is creating the right equilibrium between elasticity and viscosity, referred to as viscoelasticity. Manifestations of that can be found in everyday live, where the viscoelasticity in a tire needs to be balanced so it is still flexible and can dissipate shock-energy, yet hard enough for energy-saving operation.[1] Similarly, the cartilage in joints needs to absorb shocks while operating at low-level friction with high elasticity.[2,3]

Two such examples with a biological applicability are stiff membranes, which allow for the sliding of joints and therefore maintain their function over the lifetime of the corresponding individual (decades) and the softening of cell membranes,[2,3] for example for antimicrobial effects by dissolution in the case of bacteria (seconds).[4-6] While the first should allow for low-friction operation at high elasticity, in the second scenario energy dissipated into the membrane eventually leads to membrane destruction.

Here we address the intrinsic difference between these two types of membranes, differing in stiffness and displaying different relaxation behavior on the nanosecond time scale. The harder membranes show additional elastic modes, capillary waves, that indicate the high degree of elasticity necessary for instance in cartilage or red blood cells.

Up to now, elastic modes could not be detected as the used GINSES method was recently improved substantially by using a resonator (see supplementary material). The energy of these modes is in the order of 1 $\mu eV$.

As model systems we chose a hard phospholipidmembrane of SoyPC lipids and a $D_2O/C_{10}E_4$/decane microemulsion system representing soft surfactant membranes.

Our results help to explain properties observed for many membranes in nature, where hard membranes lubricate joints or stay intact as red blood particles in tiniest capillaries, both with extremely long lifetimes. Contrarily, softened membranes can be destroyed easily under little shear stress within seconds.

[*] Corresponding Author: s.jaksch@fz-juelich.de

# Theory

The theoretical grounds were laid by Gompper and Hennes, who motivated a relaxation function of the type:[7]

$$\frac{S(Q,\tau)}{S(Q,0)} = (1-A)\exp\left(-(\Gamma_1\tau)^\beta\right) + A\exp\left(-\Gamma_2\tau\right)\cos(\omega\tau) \qquad \text{Eq. 1}$$

The first term describes the lateral modes of single membranes at smallest energies, while the second term describes collective excitations with considerable elasticity at an energy $E_{cap} = \hbar\omega$. We attribute this energy to capillary waves (see fig. 1 c) that are collective motions of densely packed membranes. The amplitude $A$ describes the balance between the two different modes. The experimental stretch exponent $\beta$ of the single membrane modes is placed between 2/3 for over-damped membranes, 1 for a random walk of excitations, and 2 for ballistic diffusion.[8] The relaxation rates $\Gamma_1$ and $\Gamma_2$ of the two modes refer to the energy dissipation in uncorrelated undulations and damped capillary waves respectively. This is discussed in the supplementary materials. As visible in Fig. 1 a), there are clear indications for $cos(\omega\tau)$-Oscillations in the spectrum of the harder SoyPC membranes. Compared to the original theory the still strong zero energy mode was added,[7] as experimentally supported by Rheinstädter et al.[9] in the case of membrane-internal lipid motions.

Basis for this theory is the strong coupling between the fluctuating concentration field $\Phi(r)$ and the flow field $j(r)$. Taking only these two fields into account and making first approximations leads to an expression as in Eq. 1. We analyzed the formula in parallel and came to two essential conclusions that (1) the undulations in the second term of Eq.1 only occur for high bending moduli and (2) that they only occur if the motions of the membranes are strongly coupled.

# Results

The experiments whose data are presented here were performed at the NSE instrument of the Jülich Centre for Neutron Science (JCNS) located at the Spallation Neutron Source (SNS), Oak Ridge National Lab, Oak Ridge(TN), USA,[10] and at the J-NSE spectrometer at the MLZ in Garching, Germany.[11]

## SoyPC

For measurements on the SoyPC a scattering angle of $Q = 0.142$ $Å^{-1}$ was chosen. This complies well with the requirement of being as far as possible from structural features in the sample as can be seen in the reflectometry data of our previous publication,[12] to avoid a slowing of dynamics due to de Gennes narrowing.[13] Measurements were performed at the SNS.

Before starting the measurement the sample cell was completely filled with $D_2O$ and kept at 35 °C (set temperature at the Julabo thermostat) to be close to physiological conditions.

The resulting data can be seen in fig. 1 a) and were fitted using eq. 1. Results of this fit can be seen in table 1. The essential features of the relaxation curve are oscillations of the frequency ω that decay with a dampening according to $\Gamma_2$. Not only the oscillations display a high degree of elasticity, but also the low energy relaxations show a high stretching exponent of 1.6 close to perfect elasticity.

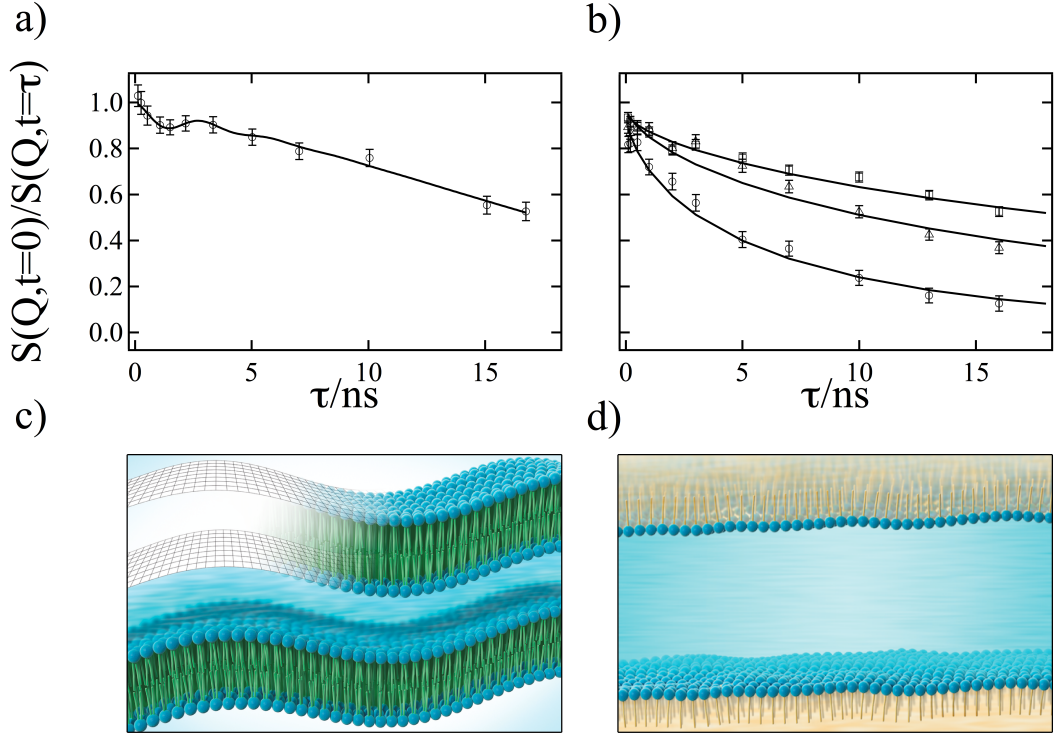

*Figure 1. GINSES data from the a) SoyPC and b) microemulsion sample as well as sketches of the corresponding systems (c and d). Lines are fits of eq. 1. Q-values in b) are Q=0.08 (squares), 0.11 (triangles) and 0.15 Å$^{-1}$ respectively. The results of these fits are shown in table 1.*

## Microemulsion

The overdamped membrane modes of microemulsions have previously been discussed for bulk systems[14] and in the presence of solid surfaces.[15] The dispersion relation for lamellar systems (with and without hard walls) introduces a $k^2$ region, where the neighbouring membranes (and wall) interact, and a $k^3$ region, where the membrane fluctuations are only embedded in the viscous medium without sensing neighbouring membranes. Here $k$ denotes the wavenumber and is not to be confused with the bending modulus $\kappa$. The first situation leads to an acceleration of the modes, while still the character of the fluctuations is due to independent membranes as long as the coupling between the membranes and the flow field is weak. The resulting relaxation times compare well with existing observations, the bending rigidity obtained from the fits to the data in Fig. 1 b) is ~1 $k_BT$. Results of fit of eq. 1 to the data are shown in table 1.

*Table 1. Resulting parameters from fit of Eq. 1 to data shown in Fig. 1 a and b.*

| Parameter | SoyPC $Q=0.142$ $Å^{-1}$ | ME $Q=0.08$ $Å^{-1}$ | ME $Q=0.11$ $Å^{-1}$ | ME $Q=0.15$ $Å^{-1}$ |
|---|---|---|---|---|
| Amplitude $A$ | 0.068±0.023 | 0 | 0 | 0 |
| Frequency $\omega$ ($ns^{-1}$) | 2.1±0.5 | N/A | N/A | N/A |
| $\Gamma_1$ (sheet) ($ns^{-1}$) and $\tau$ ($ns$) | 0.043±0.004  23.3±2.2 | 0.026±0.001  38.5±1.5 | 0.052±0.003  19.2±1.1 | 0.161±0.012  6.2±0.5 |
| Stretching exponent $\beta$ | 1.6±0.3 | 0.66 (fixed) | 0.66 (fixed) | 0.66 (fixed) |
| $\Gamma_2$ (dampening) ($ns^{-1}$) and $\tau$ ($ns$) | 0.5±0.6  2±2.4 | N/A | N/A | N/A |

# Discussion

### SoyPC

The low energy excitations display a stretching exponent $\beta$ of 1.6 between diffusion ($\beta = 1$) and ballistic diffusion ($\beta = 2$).[8] This indicates that undulations excited by thermal fluctuations or external forces will travel relatively far without strong damping. Their lifetime $\tau = 23$ $ns$ is connected to the decay rate $\Gamma_1$, which is proportional to the line width in energy space in a backscattering experiment, and through the dispersion relation to a distance of travel of 680 $nm$, considering the observed $Q$-value in inverse space ($\lambda \approx Q^{-1} = 7$ $nm$). Thus, these excitations can transport shocks over relatively large distances. We postulate these excitations exist independently for every sheet, and are to be understood as long wavelength limit ($\lambda >> Q^{-1} = 7$ $nm$) of uncorrelated undulations.

The capillary waves of wavelength $\lambda \approx Q^{-1} = 7$ $nm$ are observed as oscillations in the relaxation curve. Their energy is $E_{cap} = \hbar \omega = 1.4$ $\mu eV$, and their speed of travel is $v = 2\omega Q = 29$ $ms^{-1}$. Their lifetime is 2 $ns$ is connected to a distance of travel of 58 $nm$. To our understanding, these short wavelength undulations are continuously excited by the long wavelength undulations and serve as controlled dispersion of energy in the case of shock waves. So while the long wavelength excitations are transported over long distances, the short wavelength excitations, i.e. capillary waves quickly decay and serve for controlled damping without destruction of the bilayer. The capillary waves are a result of collective motions of several sheets (~3) that oscillate in phase.

Applying the theory of Gompper and Hennes, as laid out in the supplementary material, to the results obtained from the SoyPC measurements, we can predict several magnitudes from the simple relaxation rate of the long wavelength undulations $\Gamma_1$. The relative amplitude of the capillary waves with respect to the long wavelength undulations $A$ is 0.068 / 0.037 (measured/predicted), the frequency is 2.1 / 1.5 $ns^{-1}$ and the relaxation time is 0.5 / 0.5 $ns^{-1}$. All in all, realizing that the simple theory assumes that the bilayers are infinitely thin sheets of material in a liquid, the agreement is extremely good. This indicates, that strong coupling between the membranes is a prerequisite for observing uncorrelated long wavelength undulations and capillary waves. This coupling applies to the flow field as well as the concentration field. Long wavelength undulations are a result of many modes with relatively small amplitudes, while the capillary waves are close to the resonance of the coupling of several membranes. Their higher frequency gives rise to a faster dispersion of energy to the vicinity (i.e. liquid). All in all, this scenario is a result of rigid membranes at short distance. We put this forward as an example of visco-elasticity where the behaviour of the overall systems depends strongly on the balance between elastic and viscous properties. Here the main property if the system is elasticity. We would like to note that we neglected the hard wall as an especially hard neighbour for the first membrane due to the other already hard membranes. Then, the different theories for membranes in the neighbourhood of other planar objects become virtually identical.

### Microemulsion

The microemulsion is an example of membranes with lower bending rigidity (1 $k_BT$), and therefore with much less coupling and vanishingly small capillary waves. We extended the applied theory of Zilman and Granek[16] (applied already by Mihailescu et al.[14]) to membranes in the vicinity of a hard wall. Here, the much different rigidity of the planar wall compared to the other membranes has an impact for the first membrane: The vicinity of the interface at distance $d$ has an influence on the undulation dispersion relation for long wavelengths[15] and modifies the relaxation rate of the correlation function $S(q,t)$ depending on $d$. In summary, we observe only long wavelength undulation modes that are classically over-damped. Thus, microemulsions are highly viscous with very little elasticity.

## Summary and Conclusion

Here we have shown that the capillary waves of membranes only occur by strong coupling. We assume this is at least part of the mechanism how in nature lipid membranes can be tuned between lifetimes of $2.5 \times 10^9$ $s$ (= 80 years) in case of lipid layers providing lubrication in mammalian joints[2,3], to a few seconds in the case of a lipid layer diluted by a hygienic surfactant[5,6] or alcohol[4] as a cosurfactant. In the next paragraphs this reasoning will be elaborated in more detail.

Theoretically this can be explained by comparing the approaches of Romanov[17] and Gompper[7]. While in the approach by Romanov, coupling between the lamellae is not considered a priori, Gompper includes a coupling $B$ between the lamellae. The result is, that in the case of Romanov until very high bending moduli ($\kappa > 5000$ $k_BT$) the long wavelength dispersive modes are mainly visible and little elasticity is observed in the form of capillary modes. In the case of Gompper the capillary modes become important at bending rigidities of ca. $\kappa = 3$ $k_BT$ and thus can be observed for virtually all in vivo lipid membranes where we typically find $\kappa = 10$ to $40$ $k_BT$.

At these bending rigidities, the long wavelength modes become less dispersive and travel over long distances. The energy is dispersed through the capillary waves in a controlled manner. While the long wavelength modes appear to be rather individual for each membrane, the capillary waves are collective in-phase motions of several membranes. Contrarily, peristaltic modes would require much higher energies for excitation and the transport of liquid between the membranes would serve for much higher damping as can be seen when using the approach of Romanov.[17]

The modes, which are visible here, may seem similar to those seen by Rheinstädter et al[9,18-20], especially in terms of the mode structure. However, the latter ones appear as internal modes of the phospholipid membranes on spatial dimensions of several molecules and as a diffusion phenomenon. Here, we limit ourselves to the description of undulation and capillary membrane modes. Thus, the modes described here are of a very different nature to the previously described ones.

After we now have seen that stiffer membranes exhibit an additional mode we can assume that with this mode the membrane can absorb energy. This absorption of energy makes the membrane more stable against mechanical or chemical stress, which is why these membranes have a lifetime orders of magnitude larger than soft membranes with a bending modulus near unity. Making the assumption that this energy absorption scales exponentially with the difference in the bending modulus, as in an Arrhenius activated process, we can check this for the investigated membranes with $\kappa_{hard}$ = 20 $k_BT$ and $\kappa_{soft}$ = 1 $k_BT$, which leads to relative time difference of $\tau_{rel} \sim \exp(20)/\exp(1) \approx 1.8 \times 10^8$. If we assume a membrane with a bending modulus of unity dissolves within seconds, even this simple approximation allows for a lifetime of years for the stiffer membranes. Considering also the energy density of the states for both the membrane relaxation associated with $\Gamma_1$ and $\Gamma_2$ we see that by introducing the amplitude of the undulation mode $A$ allows for the dispersion of energy from the $\Gamma_1$ to the $\Gamma_2$ mode. There, the energy is then damped away and thus absorbed in the membrane. If there is no capillary wave mode, the complete energy is kept in the $\Gamma_1$ mode, builds up over time, and finally destroys the membrane.

We therefore attribute the stability of natural lipid bilayers that exhibit bending rigidities of $\kappa$ = 10 to 40 $k_BT$ to the strong coupling, which occurs far below the expectations of $\kappa$ = 5000 $k_BT$ without coupling. Already then, the lifetimes exceed years, contrarily to hygienic applications where soaps and cosurfactants (alcohol) lower the bending rigidity considerably, and thus reduce the lifetime to the range of seconds.

# On the difference between stiff and soft membranes: Capillary Waves

*Supplementary material*

Sebastian Jaksch[1,*], Olaf Holderer[1], Michael Ohl[2], Henrich Frielinghaus[1]

[1] Forschungszentrum Jülich GmbH, JCNS at Heinz Maier-Leibnitz Zentrum, Lichtenberstraße 1, 85747 Garching, Germany

[2] Forschungszentrum Jülich GmbH, JCNS at SNS-Oak Ridge National Laboratory (ORNL), 1 Bethel Valley Road, Oak Ridge, TN 37831, USA

## Method

Grazing-Incidence Neutron Spin Echo Spectroscopy (GINSES) is a spectroscopic method to analyse dynamics in membrane systems and combines Grazing-Incidence Small-angle Neutron Scattering (GISANS) and Neutron-Spin Echo (NSE) measurements. An image of the used geometry can be seen in Fig. 1.

In such a setup is, similarly to reflectometry, the Q-vector is perpendicular to the surface, so only dynamics in this direction can be seen and the signal is not perturbed by isotropic or in plane dynamics. A certain distance $d$ in real-space can be selected by choosing the appropriate $Q$-vector with $Q = 2 \times \pi/d$. This allows for a setup where movements in the nanosecond range on the scale of a few angstroms to several nanometres perpendicular to the surface can be investigated.

## Material

### Si Block Resonator

The silicon block resonator is a polished Si-block, where the neutrons scatter at the sample on the surface. To increase the probability of interaction between the evanescent wave and the sample a threefold double-layer system of platinum and titanium (with a repeat distance of ca. 50 *nm*) provides a waveguide, where a standing wave is created that can directly interact with the sample. This increases the probability of interaction at least one order of magnitude. The concept was inspired by the work of Nesnidal and Walker.[2]

### SoyPC Membranes

The taut membrane of SoyPC was solution casted from an isopropanol solution onto a polished silicon block and subsequently dried in vacuum. The Si-block had previously been treated with a Hellmanex solution (16 *mL* Hellmanex in 600 *mL* dI water) to achieve a clean and hydrophilic surface. This treatment was chosen in order to protect the more sensitive surface of the Si-block resonator instead of the more aggressive RCA cleaning process. The rest of the preparation replicated the procedure in our previous publication and resulted in a film of approximately 1 mm thickness.[1]

---

[*] Corresponding Author: s.jaksch@fz-juelich.de

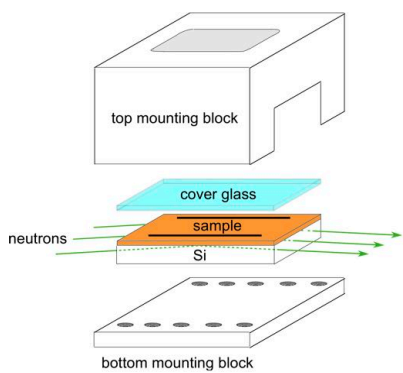

*Fig. 1. Geometry of the sample and the GINSES experiment. The rest of the NSE setup is not changed.*

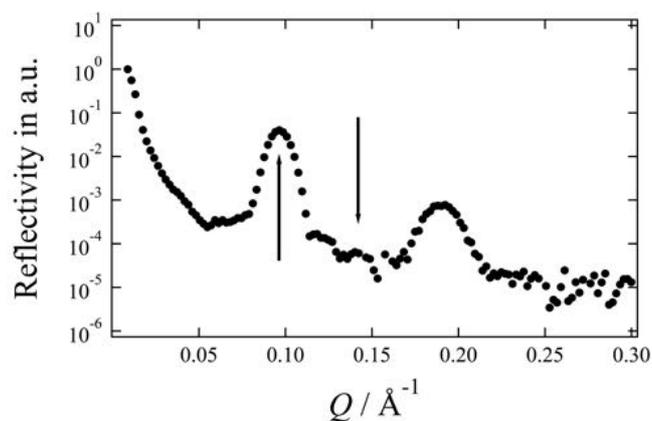

*Fig. 2. Reflectometry data of a SoyPC sample from one of our previous publications.[1] The right arrow marks the position of $Q=0.142$ Å$^{-1}$ whereas the left one points to the maximum where de Gennes narrowing occurs at $Q=0.1$ Å$^{-1}$.*

## Microemulsion

Cleaning of the Si-block followed the same protocol as in the previous case. All other preparation procedures followed the protocol as described in one of our previous publications.[3]

# Choice of the correct Q-vector

Due to the de Gennes narrowing[4] spectroscopic measurements need to be performed at a *Q*-value far from any structural peaks in the reflectometry data.

In Fig. 2 reflectometry data of a SoyPC sample can be seen. We marked the maximum position of the first peak and the measurement position. The same considerations were made for the microemulsion sample.

In both cases the choice of the *Q*-value was done a priori before the measurements. This was necessary as a scanning of several *Q*-values for an optimal position where the amplitude *A* of the capillary wave is at a maximum is not feasible with a measurement time of a day per *Q*-value.

# Theory

The amplitude $A$ of the undulations is connected to the scattering intensities $S(Q,\tau=0)$ according to:[5]

$$A \approx f(B) \cdot \left(\frac{k_0}{Q}\right)^\alpha \cdot \frac{S(Q,0)}{S(k_0,0)} \cdot \left(\frac{\xi}{d}\right)^2 \propto \left(\frac{\kappa}{k_B T}\right)^2 \qquad \text{Eq. 1}$$

The wave vector of the lamellar structure is given by $k_0=2\pi/d$ with $d$ being the lamellar thickness, and $Q = 4\times\pi \sin\vartheta/\lambda$ with $\vartheta$ as the scattering angle and the wavelength $\lambda$ is the experimental scattering vector. The exponent $\alpha$ results from scaling of the relaxation rates, for instance $\Gamma_1$, and lies between 2 and 3 for confined and free membranes. The factor is $f(B) \approx 3.0$, and becomes smaller for weaker coupling $B$ as in the case of microemulsions, and so it is plausible that not only the weak ratio of the scattering intensities of ca. 0.001 contributes to practically non-existing capillary waves for microemulsions.[6] $\xi$ is a correlation length, over which several membranes interact and $d$ is the thickness of a single membrane so the last factor is the number of membranes which is still correlated with one another. In the case of SoyPC this is usually a value of around 10.

Another important derivation for the frequency $\omega$ and for the relaxation rate $\Gamma_2$ is:

$$\omega \approx \Gamma_1\left(1-\frac{Q^2}{k_0^2}\right)^2 \frac{64}{10\pi\sqrt{3}} \frac{\kappa}{k_B T} \quad \text{and} \quad \Gamma_2 = 0.35\omega \qquad \text{Eq. 2}$$

The energy $E_{cap} = \hbar\omega$ is dependent on the relaxation rate at zero energies $\Gamma_1$ and the bending modulus of the membrane $\kappa \approx 20\,k_B T$ (for SoyPC),[7] and takes the calculated value of $\omega = 1.5\,ns^{-1}$, while experimentally we find $\omega = 2\,ns^{-1}$. The predicted rate is $\Gamma_2 = 0.5\,ns^{-1}$ and we find $\Gamma_2 = 0.5\,ns^{-1}$ experimentally. For the central coupling constant $B$ we find:

$$B \approx \frac{g_0^2}{3\pi\sqrt{3}} \frac{4096}{75} \Gamma_\Phi^2 \left(\frac{Q}{k_0}\right)^\alpha \cdot \frac{S(k_0,0)}{S(Q,0)} \cdot \left(\frac{\kappa}{k_B T}\right)^2 \qquad \text{Eq. 3}$$

Where $g_0 = 1$ is the elementary coupling constant between the concentration and flow field, and $\Gamma_\Phi = \Gamma_1(Q/Q_0)^{-\alpha}$ the elementary relaxation rate according to the elementary length scale $Q_0^{-1} = 2k_0^{-1}$. We can state that close to the correlation peak, the coupling is much weaker than at higher $Q$, where the intensities are lower. So, apart from moving away from the de Gennes narrowing[4] at the correlation peak to higher $Q$, where membrane undulations become visible, the appearance of capillary waves becomes also more likely, while here the amplitude of the capillary waves gets lower. The probability of observing capillary waves, and therefore making the system more elastic, is increasing by large bending moduli $\kappa \approx 20\,k_B T$, that has an impact on the coupling constant $B$ and the correlation peak height $S(k_0,0)$. This can also be seen by the quadratic dependence of the coupling on the bending modulus $\kappa$. The experimental scattering vector $Q$ then has to be chosen in such a way that neither the amplitude $A$ is suppressed at high $Q$-values nor the coupling $B$ vanished as very low $Q$-values. While we believe that the coupling is the central component in the concept of capillary waves at physically reasonable bending rigidities, much higher bending rigidities ($\kappa \approx 5800\,k_B T$) in other theories without coupling ($g_0 = 0$) will also lead to capillary waves.[8] This approach however we reject here, due to the unphysical values of the bending rigidity.

## Considerations for derivation

The original renormalization theory of Gompper and Hennes[5] is briefly summarized in the following. The van Hoove correlation function of the capillary waves that is accessible by neutron spin echo measurements in the time domain, reads in the energy domain:

$$S(Q,\omega) \approx 2 \cdot S(Q, \tau=0) \cdot \frac{C + \frac{B}{\hat{A}}}{\left| -i\omega + C + \frac{B}{-i\omega + \hat{A}} \right|^2} \qquad \text{Eq. 4}$$

with the coefficients:

$$\hat{A} = \Gamma_\Phi k_0^\alpha S^{-1}(k_0, 0) + \Gamma_T (Q - k_0)^2 \qquad \text{Eq. 5}$$

$$B \approx \frac{g_0^2}{3\pi\sqrt{3}} \Gamma_\Phi^2 (\xi k_0)^2 \left(\frac{Q}{k_0}\right)^\alpha \cdot \frac{S(k_0, 0)}{S(Q, 0)} \qquad \text{Eq. 6}$$

and

$$C = \Gamma_\Phi Q^\alpha S^{-1}(Q, 0) \qquad \text{Eq. 7}$$

The correlation function in the spin-echo time domain results from a Fourier transform of $S(Q,\omega)$. We identify a critical coupling constant $B = \hat{A}^2 + C^2/2 \approx C^2/2$ (we are critical, since $B \approx 3.2$ to $19 > C^2/2 \approx 1.17$), close to which we derive a relative amplitude $A$ according to eq. 1, and a frequency $\omega$ according to eq. 2. For smaller coupling constants the amplitude and energy can be smaller. For the dimensionless scattering intensity, we took:

$$S(Q,0) = \frac{k_0 \xi}{2\pi} \left( \left(1 + (k_0\xi)^{-2}\right)^2 - 2\left(1 - (k_0\xi)^{-2}\right)(Q/k_0)^2 + (Q/k_0)^4 \right)$$

$$\rightarrow \frac{64}{10\pi\sqrt{3}} \frac{\kappa}{k_B T} \left(1 - \frac{Q^2}{k_0^2}\right)^2 \qquad \text{Eq. 8}$$

in the limit of large bending rigidities $\kappa = (k_0\xi) \times 5\sqrt{3/64}$. For the intensity ratio $S(Q,0)/S(k_0,0)$, one better takes the experimental one, which we either know from small angle neutron scattering or neutron reflectivity. For reflectivity measurements, one has to add another scaling factor, according to $S(Q,0)/S(k_0,0) = (Q/k_0)^2 R(Q)/R(k_0)$.